\documentclass[aps,secnumarabic,nobalancelastpage,amsmath,amssymb,
nofootinbib,preprint]{revtex4}
\usepackage{hyperref}
\usepackage{graphics}      % standard graphics specifications
\usepackage{graphicx}      % alternative graphics specifications
\usepackage{longtable}     % helps with long table options
\usepackage{url}           % for on-line citations
\usepackage{bm}          % special 'bold-math' package
\usepackage{xcolor}
\newcommand{\beq}{\begin{equation}}
\newcommand{\eeq}{\end{equation}}
\newcommand{\bea}{\begin{eqnarray}}
\newcommand{\eea}{\end{eqnarray}}

\begin{document}

\title{Casimir Effect in the Rainbow Einstein's Universe}

\author{V. B. Bezerra\footnote{E-mail:valdir@fisica.ufpb.br}, H. F. Mota\footnote{E-mail:hmota@uaf.ufcg.br}, C. R. Muniz\footnote{E-mail:celio.muniz@uece.br}}
\affiliation{${}^{*}$Departamento de F\'isica, Universidade Federal da Para\'iba, 58059-900, Caixa Postal 5008, Jo\~ao Pessoa, PB, Brazil.\\ ${}^{\dagger}$\!Departamento de F\'sica, Universidade Federal de Campina Grande, 58429-900 , Caixa Postal 10071, Campina Grande, PB, Brazil.\\${}^{\ddag}$\!Universidade Estadual do Cear\'a, Faculdade de Educa\c c\~ao, Ci\^encias e Letras do Sert\~ao Central, 63900-000, Quixad\'a, CE, Brazil.
}

%\ead{marcony.cunha@uece.br}\address{}

%\date{}

\begin{abstract}
In the present paper we investigate the effects caused by the modification of the dispersion relation obtained by solving the Klein-Gordon equation in the closed Einstein's universe in the context of rainbow's gravity models. Thus, we analyse how the quantum vacuum fluctuations of the scalar field are modified when compared with the results obtained in the usual General Relativity scenario. The regularization, and consequently the renormalization, of the vacuum energy is performed adopting the Epstein-Hurwitz and Riemann's zeta functions.

\vspace{0.75cm}
\end{abstract}


\maketitle

\section{Introduction}

Over the past several years much effort has been made toward the acquisition of a complete theory of quantum gravity, or even toward the attempt of understanding some aspects of this expected theory. The most direct and logical way of trying to make compatible General Relativity (GR), a geometrical theory of gravity, with quantum mechanics, is within the realm of quantum field theory. This path, however, has shown to have serious problems related to the non-renormalizability of the quantum field version of gravity and, as a consequence, plenty of proposals for modification of GR has arisen so as to try to circumvent this incompatibility problem. The so called rainbow's gravity \cite{Magueijo:2002am,Magueijo:2002xx}, considered as a generalization of doubly special relativity models, is an example of such proposals.

The rainbow's gravity models, which are semi-classical proposals to investigate quantum gravity phenomena, have as fundamental principle that besides the invariant velocity of light (at low energies) there also exists an invariant energy scale set as the Planck energy (see for instance \cite{AmelinoCamelia:2000mn,Magueijo:2001cr, Galan:2004st}). In this framework, the nonlinear representation of the Lorentz transformations in momentum space leads to an energy-dependent spacetime and, consequently, to a modification of the dispersion relation \cite{AmelinoCamelia:1997gz}. This means that a probe particle will feel the spacetime differently, depending on the value of its energy and, thus, instead of a fixed spacetime geometry there will be in fact a `running geometry'. One important motivation of this new semi-classical approach is the unexplained high-energy phenomena such as the observed ultra high-energy cosmic rays, whose source is still unknown, and that suggests a modification of the dispersion relation. This opens up new possibilities for theoretical developments, mainly in areas such as astrophysics and cosmology \cite{Leiva:2008fd,Li:2008gs,Ali:2014xqa,Khodadi:2016bcx,Awad:2013nxa,Bezerra:2017hrb}.

In the cosmological realm, for instance, the rainbow's gravity acquires crucial importance since it points to the possibility of avoiding the initial singularity by means of, e.g., bouncing universe solutions \cite{Awad:2013nxa, Majumder:2013ypa} or static universes \cite{Hendi:2017vgo}. Static universes, like the Einstein universe \cite{Ford:1975su,Ford:1976fn,Herdeiro:2005zj,nariai1978quantized,grib1980particle}, with geometry having a positive constant spatial curvature are often of interest since they address issues concerning the initial singularity and cosmic horizon \cite{Ellis:2002we,Ellis:2003qz}. Another scenario where closed static universes play a central role is the one described by emergent universe models which place the Einstein Universe as the appropriate geometry to characterize the stage of the universe preceding the inflationary epoch and during which quantum vacuum fluctuations may provide robust contributions to determine parameters from these models \cite{Ellis:2002we,Ellis:2003qz}.

The Casimir effect is a phenomenon that comes about as a consequence of quantum vacuum oscillations of quantum-relativistic fields, and was originally considered as stemming from the modifications, in Minkowski spacetime, of the quantum vacuum oscillations of the electromagnetic field due to the presence of material boundaries, at zero temperature \cite{Casimir:1948dh}. The present status of the phenomenon is that the effect can also occur considering other fields (such as the scalar and spinor fields), and also due to nontrivial topologies associated with different geometries that may describe the spacetime, like for instance the static Einstein universe (for reviews, see \cite{milton2001casimir,bordag2009advances}). The Casimir effect arising as a consequence of the nontrivial topology of the static Einstein and Friedman universes has been investigated in a variety of works \cite{Bezerra:2011nc,Bezerra:2011zz,Zhuk:1996xc,Bezerra:2014pza,Mota:2015ppk,Bezerra:2016qof}, in which the role of the quantum vacuum energy in the primordial universe is highlighted. Here we are interested in studying the modification of the quantum vacuum energy caused by the nontrivial topology of the Einstein universe, in the context of the rainbow's gravity models.

The paper is organized as follows: In section 2 we make a brief review of the rainbow's gravity framework. In section 3 we compute the quantum vacuum energy caused by the Einstein rainbow's universe and, finally, in section 4, we present our final remarks.

%%%%%%%%%%%%%%%%%%%%%%%%%%%%%%%%%%%%%
\section{Rainbow's gravity framework}
\label{sec2}
%%%%%%%%%%%%%%%%%%%%%%%%%%%%%%%%%%%%%
In the framework of rainbow's gravity, also known as doubly General Relativity, probe particles can energetically influence the spacetime background, making it energy-dependent so that a variety of metrics is possible. This assumption, of course, has the implication that at high energy scales the sensitivity of the metric to the energy of the probe particles leads to a modified dispersion relation \cite{Magueijo:2002am,Magueijo:2002xx}, i.e.,
\begin{equation}
E^2g_0(x)^2 - p^2g_1(x)^2 = m^2c^4,
\label{MDR}
\end{equation}
where $x=\frac{E}{E_P}$ is the ratio of the energy of the probe particle to the Planck energy $E_P$, and regulates the level of the mutual relation between the spacetime background and the probe particles. The functions $g_0(x)$ and $g_1(x)$ are called rainbow functions and should be chosen so that the model can be best fit to the data and explain known cosmological puzzles \cite{AmelinoCamelia:2008qg}. As the rainbow's gravity models are semi-classical approaches to understand the quantum gravity regime, the rainbow functions should be also consistent with quantum gravity theory attempts such as loop quantum gravity, non-commutative geometries and so on. Evidently, in the low energy regime one should have
\begin{equation}
\lim_{x\rightarrow 0}g_{i}(x) = 1, with i=0,1.
\end{equation}

In the context of the rainbow's gravity, the Friedmann-Robertson-Walker(FRW) spacetime, with topology $S^3\times R^1$ has been argued to be described by a line element having the general form  \cite{Awad:2013nxa, Hendi:2017vgo, Ling:2006az, Ling:2008sy}
\begin{equation}\label{FRWLE}
ds^2 = \frac{c^2dt^2}{g_0(x)^2} - \frac{a^2(t)}{g_1(x)^2}\left[\frac{dr^2}{(1-\chi r^2)} + r^2(d\theta^2 + \sin^2\theta d\phi^2)\right],
\end{equation}
where $a(t)$ is the time-dependent scale factor and $\chi$ is the constant spatial curvature that can take the values $-1,0,1$, corresponding to a spacetime with open, flat or closed spatial curvature, respectively. As usual, the spacetime coordinates take values in the following ranges: $-\infty < t < +\infty$, $r\geq 0$, $0\leq \theta\leq\pi$ and $0\leq \phi\leq 2\pi$. The line element \eqref{FRWLE} has been considered in Refs. \cite{Awad:2013nxa, Hendi:2017vgo, Ling:2006az, Ling:2008sy}
 to investigate the modifications of the Friedmann equations and all properties of the cosmological physical quantities caused by the rainbow functions, in special, the possibility of a nonsingular universe.

In the present work, we want to consider the three mostly adopted rainbow functions. The first of them is given by
\begin{equation}
g_0(x) = g_1(x) =  \frac{1}{1- x}.
\label{rainf1}
\end{equation}
This rainbow function has been considered in Refs.  \cite{Awad:2013nxa,Hendi:2017vgo,Khodadi:2016bcx} (see also references therein) and provides a constant velocity of light as well as accounts for the horizon problem. The authors of Refs.  \cite{Awad:2013nxa,Hendi:2017vgo,Khodadi:2016bcx} considered this rainbow function to investigate the effects of the gravity rainbow on the FRW universe, in special,  solutions corresponding to nonsingular universe.

The second rainbow function we would like to consider is written as
\begin{equation}
g_0(x)=1,\qquad\qquad g_1(x) =\sqrt{1-x^2}.
\label{rainf2}
\end{equation}
This rainbow function has also been considered in Refs.  \cite{Awad:2013nxa,Hendi:2017vgo,Khodadi:2016bcx} to investigate the effects of the gravity rainbow on the FRW universe.

Finally, the third rainbow function we would like to consider is given by
\begin{equation}
g_0(x)=\frac{e^x - 1}{x},\qquad\qquad g_1(x) =1.
\label{rainf3}
\end{equation}
This rainbow function was considered also in Ref.  \cite{Awad:2013nxa} and was originally proposed in \cite{AmelinoCamelia:1997gz} to explain gamma-ray burst phenomena in the universe.

By adopting the three rainbow functions written above we want to obtain closed expression for the modification of the renormalized vacuum energy in the Einstein universe, which is in fact a universe described by the line element in Eq. \eqref{FRWLE}, with a constant scale factor and with closed spatial curvature. The dispersion relation used to calculate the renormalized vacuum energy is obtained by solving the Klein-Gordon equation considering the line element \eqref{FRWLE}. This has been done in a number of previous works \cite{Ford:1975su,Ford:1976fn,Herdeiro:2005zj,nariai1978quantized,grib1980particle} in the context of the conventional FRW universe and we will only use the result to modify it according to the modified dispersion relation \eqref{MDR}. Let us then, in the next section, introduce the calculation of the renormalized vacuum energy obtained in the conventional Einstein universe.
%%%%%%%%%%%%%%%%%%%%%%%%%%%%%%%%%%%%%
\section{Quantum vacuum energy from FRW rainbow's gravity}
%%%%%%%%%%%%%%%%%%%%%%%%%%%%%%%%%%%%%
\subsection{Conventional FRW gravity}
%%%%%%%%%%%%%%%%%%%%%%%%%%%%%%%%%%%%%
%
The solution of the Klein-Gordon equation for a conformally coupled massive scalar field in the Einstein Universe, with topology $S^3\times R^1$, has been considered, for instance, in Refs \cite{Ford:1975su,Ford:1976fn,Herdeiro:2005zj,nariai1978quantized,grib1980particle}. The eigenfrequencies in this case are given in terms of the constant scale factor $a_0$ as
\begin{equation}\label{EigenEnergiesWithCoupling}
\omega_n=\frac{ c}{a_0}(n^2+\nu^2)^{\frac{1}{2}},
\end{equation}
where $\nu=\frac{a_0cm}{\hbar}$, with $m$ being the mass of the scalar particle. Note that the eigenfrequencies above are also valid for the static FRW universe as well as all the results obtained below, as argued in \cite{Bezerra:2011nc,Bezerra:2011zz}. In the conventional static FRW universe, nonetheless, there also exists additional contributions to the energy-momentum tensor due to the conformal anomaly and creation of particles \cite{Bezerra:2011nc,Bezerra:2011zz} and we should expect that this is also true in the context of the rainbow's FRW universe. Note, however, that in the present work we wish only to investigate, as a first step, the effects of the rainbow's Einstein universe on the quantum vacuum energy which we will deal with in the next section by modifying the eigenfrequencies \eqref{EigenEnergiesWithCoupling} according to  \eqref{MDR}.

The standard procedure to obtain the quantum vacuum energy is to sum the eigenfrequencies \eqref{EigenEnergiesWithCoupling} over the modes of the field as \cite{Bezerra:2011nc,Bezerra:2011zz}
\begin{eqnarray}\label{VacuumEnergyWithCoupling}
E_0&=&\sum_{n=1}^{\infty}n^2\omega_n=\frac{\hbar c}{2a_0}\sum_{n=1}^{\infty}n^2(n^2+\nu^2)^{\frac{1}{2}}\nonumber\\
&=&\frac{\hbar c}{2a_0}\left[\sum_{n=1}^{\infty}(n^2+\nu^2)^{\frac{3}{2}}-\nu^2\sum_{n=1}^{\infty}(n^2+\nu^2)^{\frac{1}{2}}\right],
\end{eqnarray}
which is a non-regularized infinity vacuum energy. In order to regularize this divergent expression we use the Epstein-Hurwitz zeta function defined as
\begin{equation}
\zeta_{EH}{(s,\nu)}=\sum_{n=1}^{\infty}(n^2+\nu^2)^{-s}
\label{EH}
\end{equation}
for $\mathrm{Re}(s)>1/2$ and $\nu^2\geq 0$. Note that the latter restriction is fully satisfied.  We can now use the representation of $\zeta_{EH}{(s,\nu)}$ that yields its analytical continuation for other values of $s$, and that is given in terms of the Macdonald function, $K_{\nu}(z)$, as follows
\begin{equation}\label{RepresentationEpsteinHurwitz}
\zeta_{EH}{(s,\nu)}=-\frac{\nu^{-2s}}{2}+\frac{\sqrt{\pi}}{2}\frac{\Gamma(s-\frac{1}{2})}{\Gamma(s)}\nu^{(1-2s)}+\frac{2\pi^s\nu^{(\frac{1}{2}-s)}}{\Gamma(s)}\sum_{n=1}^{\infty}n^{(s-\frac{1}{2})}K_{(s-\frac{1}{2})}(2\pi n \nu).
\end{equation}
Thus, in order to regularize the divergent vacuum energy \eqref{VacuumEnergyWithCoupling} we can make use of the Epstein-Hurwitz zeta function and its analytic continuation in Eq. \eqref{RepresentationEpsteinHurwitz} \cite{elizalde1994zeta}. By doing this we will still have divergent contributions, of the form $\Gamma(-2)$ and $\Gamma(-1)$, coming from the second term on the r.h.s of \eqref{RepresentationEpsteinHurwitz}. However, such terms should be subtracted since it comes from the continuum part of the sums in Eq. \eqref{VacuumEnergyWithCoupling} \cite{elizalde1994zeta, Nesterenko:1997ku}. This term can, for instance, be obtained by the substitution
\begin{equation}\label{cont}
\sum_{n=1}^{\infty}(n^2+\nu^2)^{-s}\qquad \Longrightarrow\qquad\int_{0}^{\infty}(t^2+\nu^2)^{-s}dt = \frac{\sqrt{\pi}}{2}\frac{\Gamma(s-\frac{1}{2})}{\Gamma(s)}\nu^{(1-2s)},
\end{equation}
which can be better seen through the Abel-Plana formula. Therefore, the regularized vacuum energy associated with the conformally coupled massive scalar field in the Einstein universe is found to be
\begin{equation}\label{CasimirEnergyWithmass}
E_0^{\mathrm{ren}}=\frac{\hbar c\nu^4}{a_0}\left[3\sum_{n=1}^{\infty}f_{2}(2\pi n \nu)+\sum_{n=1}^{\infty}f_{1}(2\pi n \nu)\right],
\end{equation}
where the function $f_{\nu}(y)$ is written in terms of the Macdonald function as
\begin{equation}
f_{\nu}(y) =\frac{K_{\nu}(y)}{y^{\nu}}.
\label{besself}
\end{equation}
Note that in the massless scalar field limit, that is, $m\rightarrow 0$, the result in Eq. \eqref{CasimirEnergyWithmass} reduces to the well known result \cite{Ford:1976fn}
\begin{equation}\label{CasimirEnergyWithmassless}
E_0^{\mathrm{ren}}=\frac{3\hbar c}{8\pi^4 a_0}\zeta(4) = \frac{\hbar c}{240 a_0},
\end{equation}
where $\zeta(s)$ is the Riemann zeta function, which provides $\zeta(4)=\frac{\pi^4}{90}$. We can see that the dropping of the gamma divergent terms in Eq. \eqref{VacuumEnergyWithCoupling} truly leads to the correct result for the massless scalar field case.
It is worth calling attention that the methods which apply the generalized Riemann zeta functions have been used in different contexts of both the semiclassical conventional gravity and rainbow's gravity scenarios (\cite{Faizal,Mistry} and references therein).

%%%%%%%%%%%%%%%%%%%%%%%%%%%%%%%%%%%
\subsection{FRW rainbow's gravity}
%%%%%%%%%%%%%%%%%%%%%%%%%%%%%%%%%%%
%
We want now to analyse how the regularized vacuum energies \eqref{CasimirEnergyWithmass}-\eqref{CasimirEnergyWithmassless} change in the context of rainbow's gravity. For this we will make use of the three rainbow functions presented in Sec.\ref{sec2}. Let us then start with the first one given by Eq. \eqref{rainf1}.
\begin{itemize}
\item[{\bf(1)}] {\bf Case:} $g_0(x)=g_1(x)= \frac{1}{1-\epsilon x}$
\end{itemize}

For this case we shall consider the eigenfrequencies of a massive scalar field given by Eq. \eqref{EigenEnergiesWithCoupling}. The modification of the latter is performed according to the modified dispersion relation \eqref{MDR}, which has the resulting effect of making $m\rightarrow \frac{m}{g_0}$ in \eqref{EigenEnergiesWithCoupling}. This provides
\begin{equation}\label{rainboweingen}
x_n^2 + \epsilon m_0^2x - m_0^2 - x_0^2n^2 =0,\qquad\mathrm{with}\qquad x_n = \frac{\hbar\omega_n}{E_P},\;\;\;\; m_0 = \frac{mc^2}{E_P},\;\;\;\; x _0= \frac{\hbar c}{a_0E_P}.
\end{equation}
Note that in order to keep track of the part of the vacuum energy that will be modified we have introduced the order one parameter,
$\epsilon$, in the rainbow function. Note also that, since the Planck energy $E_P$ is set as the maximum scale energy in the rainbow's gravity scenario, the range of the parameters $x_n$, $m_0$ and $x_0$ defined above is within the interval going from zero to one. Thereby, the relevant physical solution of Eq. \eqref{rainboweingen} is given by
\begin{equation}\label{sol}
x_n = -\frac{\epsilon m_0^2}{2}+x_0 [n^2 + p^2]^{\frac{1}{2}},
\end{equation}
where $p^2=\frac{\epsilon m_0^4+4m_0^2}{4x_0^2}$. Thus, if the terms with $\epsilon$ in \eqref{sol} are neglected we recover the expression for the eigenfrequencies in \eqref{EigenEnergiesWithCoupling}.

Following the same spirit as in Eq. \eqref{VacuumEnergyWithCoupling}, the regularized vacuum energy can be write in terms of the Epstein-Hurwitz zeta function as
\begin{eqnarray}\label{ZPE1}
E_0 &=&\frac{E_P}{2}\sum_{n=1}^{\infty}n^2x_n\nonumber\\
&=&-\frac{m_0^2E_P}{4}\zeta(-2)+\frac{x_0E_P}{2}[\zeta_{EH}(-3/2;p)-p^2\zeta_{EH}(-1/2;p)]\nonumber\\
&=&\frac{x_0E_P}{2}[\zeta_{EH}(-3/2;p)-p^2\zeta_{EH}(-1/2;p)],
\end{eqnarray}
where we have used the property $\zeta(-2k)=0$. Furthermore, using the analytic extension \eqref{RepresentationEpsteinHurwitz} of the Epstein-Hurwitz zeta function, after discarding the divergent parts that depend on $\Gamma(-1)$ and $\Gamma(-2)$, as we did previously, we obtain the renormalized quantum vacuum energy
\begin{eqnarray}\label{RQVE}
E_0^{\mathrm{ren}} =p^4x_0E_P\left[3\sum_{n=1}^{\infty}f_2(2\pi np)+\sum_{n=1}^{\infty}f_1(2\pi np)\right].
\end{eqnarray}
The modification introduced by the rainbow function considered in the present case is codified in the definition of $p$ below Eq. \eqref{sol}. If we neglect the term with $\epsilon$ in this definition we obtain $p=\nu$, and we recover the vacuum energy \eqref{CasimirEnergyWithmass} obtained in the previous section. Note that an important feature of the rainbow function considered here is that, in the massless limit of Eq. \eqref{RQVE}, there is no modification of the vacuum energy since the resulting expression coincides with the one in Eq. \eqref{CasimirEnergyWithmassless}. This can be seen direct from the general modified dispersion relation \eqref{MDR}. In the massless case, the r.h.s of this expression is zero and as $g_0(x)=g_1(x)$ one can factorize these functions out in such a way to have no modification.
%
%%%%%%%%%%%%%%%%%%%%%%%%%%%%%%%%%%%%%%%%%%%%%%%%%%%%%%%%%%%%%%%%%%%%%%%%%%%%%%%%%%%%%%
\begin{figure}[!htb]
\begin{center}
\includegraphics[width=0.6\textwidth]{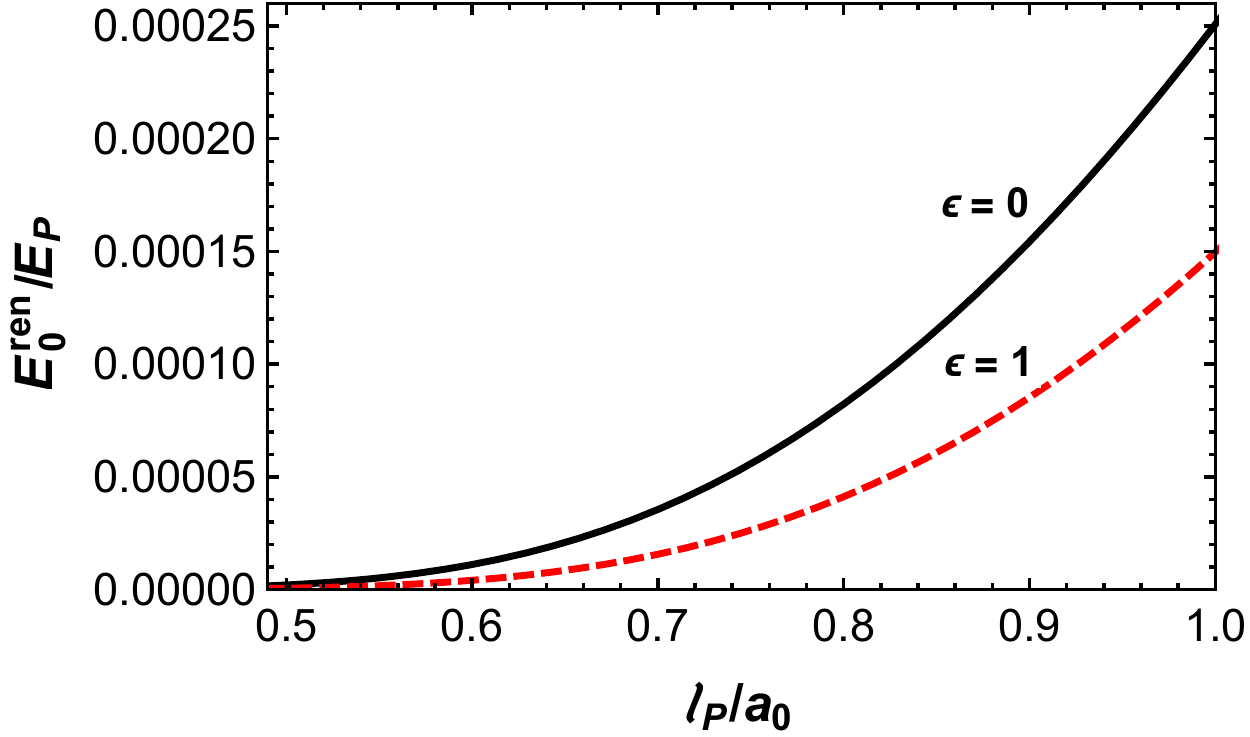}
\caption{\small{Plot of the ratio $\frac{E_0^{\mathrm{ren}}}{E_P}$ in terms of the ratio $\frac{\ell_P}{a_0}$, in the massive case, considering the rainbow functions $g_0(x)=g_1(x)= \frac{1}{1-\epsilon x}$. Note that the Planck length is given by $\ell_P=\frac{\hbar c}{E_P}$.}}
\label{f1}
\end{center}
\end{figure}
%%%%%%%%%%%%%%%%%%%%%%%%%%%%%%%%%%%%%%%%%%%%%%%%%%%%%%%%%%%%%%%%%%%%%%%%%%%%%%%%%%%%%%%%%%%
%
In Fig.\ref{f1} we have plotted the ratio of the renormalized vacuum energy \eqref{RQVE} to the Planck energy $E_P$, in terms of the ratio of the Planck length to the constant scale factor, for the massive case. We can see that, compared with the known result for $\epsilon =0$, the ratio $\frac{E_0^{\mathrm{ren}}}{E_P}$ for $\epsilon =1$ decreases. We can also see that the effects appearing as a consequence of the rainbow function \eqref{rainf1} is more apparent from $\frac{\ell_P}{a_0} \simeq 0.5$, where the Planck length is defined as $\ell_P=\frac{\hbar c}{E_P}$.
%%%%%%%%%%%%%%%%%%%%%%%%%%%%%%%%%%%%%
\begin{itemize}
\item[{\bf(2)}] {\bf Case:} $g_0(x)=1; \; g_1(x)= \sqrt{1-\epsilon x^2}$
\end{itemize}
%%%%%%%%%%%%%%%%%%%%%%%%%%%%%%%%%%%%

The second rainbow function we would like to make use is the one given by Eq. \eqref{rainf2}. The resulting effect of this rainbow function on the eigenfrequencies  \eqref{EigenEnergiesWithCoupling} is codified in the scale factor which changes according to $a_0\rightarrow \frac{a_0}{g_1(x)}$. This rainbow function is also inspired by the loop quantum gravity theory and non-commutative space models of gravity. Thus, the new modified eigenfrequencies are solutions of the equation
\begin{equation}
x_n^2 +\epsilon x_{0}^2n^2 x^2 - x_{0}^2n^2 - m_0^2=0,
\label{massive}
\end{equation}
where the parameters $x_n$, $m_0$ and $x_0$ are the same as the ones defined in Eq. \eqref{sol}. For this case, the only solution that provides positive eigenfrequencies  is given by
\begin{equation}\label{s=2}
x_n=\frac{(n^2+p^2)^{\frac{1}{2}}}{\left[1+\epsilon x_{0} ^2n^2\right]^{\frac{1}{2}}}x_{0},
\end{equation}
where $p=\frac{m_0}{x_0}$. Note that if we set $\epsilon =0$ in the above equation, we recover the eigenfrequencies given by Eq. \eqref{EigenEnergiesWithCoupling}. This is a rather complicated spectrum of eigenfrequencies to make use of the Epstein-Hurwitz zeta function in an easy way. Because of that, let us then consider the massless case, i.e., $p=0$, in order to calculate the quantum vacuum energy
\begin{eqnarray}\label{ZPE2}
E_0 &=&\frac{E_P}{2}\sum_{n=1}^{\infty}n^2x_n\nonumber\\
&=&\frac{E_Px_0}{2}\sum_{n=1}^{\infty}\frac{n^3}{\left[1+\epsilon x_{0} ^2n^2\right]^{\frac{1}{2}}}.
\end{eqnarray}
This vacuum energy is infinity and we need to regularize it in order to drop the divergent term and obtain the final renormalized vacuum energy. For this, we can use a binomial expansion so as to express the vacuum energy in terms of the Riemann's zeta function. This provides
\begin{eqnarray}\label{ZPE3}
E_0^{\mathrm{ren}}=\frac{x_0E_P}{240}+\frac{x_0E_P}{2\sqrt{\pi}}\sum_{k=1}^{\infty}\frac{\zeta(-2k-3)\Gamma\left(\frac{1}{2}+k\right)}{\Gamma(k+1)}(\sqrt{\epsilon}x_0)^{2k}.
\end{eqnarray}
Note that the first term on the r.h.s is the renormalized vacuum energy of a massless scalar field given by Eq. \eqref{CasimirEnergyWithmassless} and the second term is the correction due to the modification of the dispersion relation \eqref{EigenEnergiesWithCoupling} by the rainbow function considered in the present case. Moreover, we have numerically checked that the summation in $k$ above converges as long as $\sqrt{\epsilon} x_0\lesssim 6.7\times10^{-5}$, which corresponds to $a_0 \gtrsim  1.5\times10^{-31}\;m$ or, in other words, $a_0 \gtrsim  10^4\ell_P$. That is, the minimum value for the scale factor is bigger than the Planck length by a factor of about $10^4$, which is in accordance with the requirement that the ratio $\frac{\ell_P}{a_0}$ should be smaller or equal to one if we  consider the Planck length as the minimum invariant length as the gravity rainbow requires. We can estimate the value introduced by the correction on the r.h.s of Eq. \eqref{ZPE3} to the vacuum energy \eqref{CasimirEnergyWithmassless} by taking $\frac{\ell_P}{a_0}= 10^{-4}$, which provides the maximum value for the correction. For this value of the ratio $\frac{\ell_P}{a_0}$ we found, in units of the Planck energy, that the correction is given by $\simeq -3\times 10^{-16}$. This is very small compared to the vacuum energy \eqref{CasimirEnergyWithmassless} which, for this value of $\frac{\ell_P}{a_0}$, in units of the Planck energy, provides $\simeq 2.8\times 10^{-7}$.
%
%%%%%%%%%%%%%%%%%%%%%%%%%%%%%%%%%%%%%%%%%%
\begin{itemize}
\item[{\bf(3)}] {\bf Case:} $g_0(x)=\frac{e^{\epsilon x}-1}{\epsilon x}; \; g_1(x)= 1$
\end{itemize}
%%%%%%%%%%%%%%%%%%%%%%%%%%%%%%%%%%%%%%%%%%
%

The third and last rainbow function we want to consider now is the one given by Eq. \eqref{rainf3}. This rainbow function was proposed by Amelino-Camelia and collaborators to explain high-energy cosmic ray phenomena \cite{AmelinoCamelia:1997gz,AmelinoCamelia:2008qg}. Thus, the modification of the eigenfrequencies \eqref{EigenEnergiesWithCoupling} due to this rainbow function is written as
\begin{eqnarray}
x_n = \frac{1}{\epsilon}\ln\left(\epsilon\sqrt{m_0^2 + x_0^2n^2} +1\right) = \sum_{k=1}^{\infty}\frac{(-1)^{k+1}\epsilon^kx_0^{k}(\nu^2+n^2)^{\frac{k}{2}}}{\epsilon k},
\label{case3}
\end{eqnarray}
where $\nu=\frac{m_0}{x_0}=\frac{a_0cm}{\hbar}$ and we have used the series expansion for the logarithmic function. As previously done, the vacuum energy is obtained through
\begin{eqnarray}\label{case3.1}
E_0 &=&\frac{E_P}{2}\sum_{n=1}^{\infty}n^2x_n=\frac{E_P}{2}\sum_{n=1}^{\infty}n^2 \sum_{k=1}^{\infty}\frac{(-1)^{k+1}\epsilon^kx_0^{k}(\nu^2+n^2)^{\frac{k}{2}}}{\epsilon k}\nonumber\\
&=&\frac{E_P}{2} \sum_{k=1}^{\infty}\frac{(-1)^{k+1}\epsilon^{k-1}x_0^{k}}{ k}[\zeta_{EH}(-k/2-1;\nu)-\nu^2\zeta_{EH}(-k/2;\nu)],
\end{eqnarray}
which is given in terms of Epstein-Hurwitz zeta functions. Note that the $k=1$ term above provides the vacuum energy \eqref{VacuumEnergyWithCoupling} of the massive scalar field without the rainbow function corrections. This term was analysed in the beginning of the present section. As to the other values of $k$, the vacuum energy is also infinity and the regularization can be performed by means of the Epstein-Hurwitz function analytic extension \eqref{RepresentationEpsteinHurwitz}. By using the latter in Eq. \eqref{case3.1}, the first term cancels out and the second term diverges for odd values of $k$. These divergent terms should be subtracted as discussed earlier. On the other hand, the contribution of the even values of $k$ to the second term of Eq. \eqref{RepresentationEpsteinHurwitz} goes to zero since the gamma function in the denominator goes to infinity. This leave us only with the last term on the r.h.s of Eq. \eqref{RepresentationEpsteinHurwitz}, which provides the following renormalized finite vacuum energy:
\begin{eqnarray}\label{case3.2}
E_0^{\mathrm{ren}}&=&x_0E_P\nu^4\left[3\sum_{n=1}^{\infty}f_{2}(2\pi n \nu)+\sum_{n=1}^{\infty}f_{1}(2\pi n \nu)\right]\nonumber\\
&+&\frac{E_P\sqrt{\pi}}{2} \sum_{k=2}^{\infty}2^{\frac{k+3}{2}}\frac{(-1)^{k+1}x_0^{k}\nu^{k+3}}{ k}\sum_{n=1}^{\infty}\left[\frac{2f_{\frac{k+3}{2}}(2\pi n\nu)}{\Gamma( - \frac{k}{2}-1)} - \frac{f_{\frac{k+1}{2}}(2\pi n\nu)}{\Gamma(- \frac{k}{2})}\right],
\end{eqnarray}
where the function $f_{\mu}(x)$ was defined in Eq. \eqref{besself} and we have taken $\epsilon= 1$ (as we pointed out before, this is a order one parameter). The first term on the r.h.s of the above expression is the vacuum energy \eqref{CasimirEnergyWithmass} without the rainbow function corrections and the second term comprises the corrections. Furthermore, by using the asymptotic expression of the Macdonald function for small arguments  \cite{Abramowitz} we can obtain the Casimir vacuum energy for the massless scalar field, which is given by
\begin{eqnarray}\label{case3.2massless}
E_0^{\mathrm{ren}}&=& \frac{x_0E_P}{240} + \frac{E_P\sqrt{\pi}}{2} \sum_{k=2}^{\infty}\frac{(-1)^{k+1}x_0^{k}}{ k\pi^{k+3}}\frac{\Gamma(\frac{k+3}{2})}{\Gamma(-\frac{k}{2}-1)}\zeta(k+3).
\end{eqnarray}
Now, using the properties of both the zeta and gamma functions \cite{Abramowitz}, we are able to show that
\begin{equation}
\zeta(-k-2)=\frac{\sqrt{\pi}}{\pi^{k+3}}\frac{\Gamma(\frac{k+3}{2})}{\Gamma(-\frac{k}{2}-1)}\zeta(k+3).
\end{equation}
Taking into account this expression, we can re-write Eq. \eqref{case3.2massless} as
\begin{eqnarray}\label{case3.2massless2}
E_0^{\mathrm{ren}}&=& \frac{x_0E_P}{240}  + \frac{E_P}{2} \sum_{k=2}^{\infty}\frac{(-1)^{k+1}x_0^{k}}{ k}\zeta(-k-2),
\end{eqnarray}
which is finite for all allowed values of $x_0$ . The form of the vacuum energy for the massless case presented above is exactly the one obtained when we use the eigenfrequencies \eqref{EigenEnergiesWithCoupling} without mass from the beginning. This shows that the regularization procedure, and consequently renormalization, of Eq. \eqref{case3.1} using the Epstein-Hurwitz zeta function is consistent, and thus in agreement with what is usually done in the literature \cite{elizalde1994zeta, Nesterenko:1997ku}. By taking $x_0=10^{-2}$, for instance, the correction in the second term on the r.h.s of Eq. \eqref{case3.2massless2} is estimated to be $\simeq 6.6\times 10^{-10}E_P$, which is several orders of magnitude smaller than the first term, estimated as $\simeq 4.2\times 10^{-5}E_P$ for the value of $x_0$ considered. These values decrease $x_0$ decreases, but the vacuum energy given by the first term in \eqref{case3.2massless2} is always greater than the correction.
%
%%%%%%%%%%%%%%%%%%%%%
\section{Summary and Discussion}
%%%%%%%%%%%%%%%%%%%%
%
We have calculated the renormalized vacuum energy of a conformally coupled scalar field in the FRW spacetime with a constant scale factor and positive spatial curvature (Einstein's Universe) in the scenario of the rainbow's gravity, considering three different rainbow functions given by Eqs. \eqref{rainf1}, \eqref{rainf2} and \eqref{rainf3}. The rainbow's gravity approach presupposes a metric depending on the energy of the probe particle and, as a consequence, a modified dispersion relation with the general form given by Eq. \eqref{MDR}. Thus, the quantum gravity effects are more pronounced as the particle's energy approaches the Planck scale. The calculation was made at zero temperature and the regularization technique was based on the use of Epstein-Hurwitz and Riemann's zeta functions.

The first rainbow function considered, i.e., Eq. \eqref{rainf1}, provided a modified dispersion relation given by Eq. \eqref{sol}, leading to a renormalized vacuum energy \eqref{RQVE}. In Fig.\ref{f1}, this renormalized vacuum energy was plotted, in units of the Planck energy, in terms of the ratio $\frac{\ell_P}{a_0}$ and shows the modification of the vacuum energy in the context of the rainbow's gravity when compared with the vacuum energy obtained in the Einstein universe. In this plot, we showed that the higher the ratio, the greater the difference between the Casimir energies with and without the corrections due to the rainbow function considered. We also pointed out that in the massless limit both vacuum energies coincide, that is, the rainbow's gravity does not modify the vacuum energy obtained considering the Einstein universe from General Relativity.

The second rainbow function considered in Eq. \eqref{rainf2} provided the modified dispertion relation given by Eq. \eqref{s=2}. For simplicity, we only took into account the massless limit. In this case, it was shown that the renormalized vacuum energy is found to be the one in Eq. \eqref{ZPE3}, which converges only for $x_0\lesssim 6.7\times10^{-5}$. Taking into account the maximum allowed value $x_0$, we estimate that the correction to the vacuum energy which is very small.

Finally, in the last studied case, where we considered the rainbow function in Eq. \eqref{rainf3}, we found that the resulting vacuum energy is given by Eq. \eqref{case3.2} in the massive case and Eq. \eqref{case3.2massless} in the massless case. These vacuum energies were a consequence of the modified dispersion relation in \eqref{case3}. Again we estimated that, for instance, considering $x_0=10^{-2}$ in the massless case, the correction to the vacuum energy is many orders of magnitude smaller.

In general, the smaller vacuum energy associated with the rainbow gravity seems to be a characteristic of this theory due to the fact that part of the field energy is spent in deforming the proper spacetime where it is (a kind of backreaction effect) \cite{Bezerra:2017hrb}. On the other hand, the monotonicity of the corrections - the fact that the Casimir energy grows continuously with the diminishment of the Universe scale factor - can be explained in a simple way by the growing of the vacuum fluctuations associated with the reduction of the corresponding spatial volume.

As a future perspective, we intend to analyse finite temperature corrections to the vacuum energies considered here as well as rainbow's gravity corrections to the vacuum energy considering other fields, as for instance the spinor and electromagnetic fields.
%
%%%%%%%%%%%%%%%%%%%%
\section*{Acknowledgments}
%%%%%%%%%%%%%%%%%%%%
%
We would like to thank Eug\^enio R. B. de Mello for useful discussions. V.B.B and C.R.M are supported by the Brazilian agency CNPq (Conselho Nacional de Desenvolvimento Cient\'{i}fico e Tecnol\'{o}gico). H.F.M is supported by the Brazilian agency CAPES (Coordena\c{c}\~ao de Aperfei\c{c}oamento de Pessoal de N\'ivel Superior).
%

%\bibliography{refs}

 \end{document}